\documentclass[prl,twocolumn,showpacs]{revtex4}
\usepackage{hyperref}
\usepackage{amssymb, amsmath}
\usepackage[dvips]{graphicx}
\usepackage{latexsym}
\usepackage{verbatim}

\newcommand{\de}{\mathrm{d}}

\begin{document}
\title{Effect of magnetic impurities on energy exchange between electrons}
\date{\today}
\author{B. Huard}
\author{A. Anthore}
\altaffiliation{Present address : Mat\'eriaux et Ph\'enom\`enes
Quantiques, Universit\'e Paris 7, 2 place Jussieu, 75251 Paris,
France.}
\author{Norman O. Birge}
\altaffiliation{Permanent address : Department of Physics and
Astronomy, Michigan State University, East Lansing, MI 48824}
\author{H. Pothier}
\email[Corresponding author: ]{hpothier@cea.fr}
\author{D. Esteve}
\affiliation{Quantronics Group, Service de Physique de l'\'Etat
Condens\'{e}, DRECAM, CEA-Saclay, 91191 Gif-sur-Yvette, France}

\begin{abstract}
In order to probe quantitatively the effect of Kondo impurities on
energy exchange between electrons in metals, we have compared
measurements on two silver wires, with dilute magnetic impurities
(manganese) introduced in one of them. The measurement of the
temperature dependence of the electron phase coherence time on the
wires provides an independent determination of the impurity
concentration. Quantitative agreement on the energy exchange rate is
found with a theory by G\"oppert \emph{et al.} that accounts for
Kondo scattering of electrons on spin-1/2 impurities.
\end{abstract}  \maketitle

In diffusive metals, it is expected that the dominant inelastic
electron scattering process at low temperature is the Coulomb
interaction~\cite{AAK,Gilles1}, leading to a power law increase of
the electron phase coherence time $\tau_{\varphi}$ with decreasing
temperature $T$.  However, in the presence of a small concentration
of magnetic impurities with low Kondo temperature, $\tau_{\varphi}$
can be limited by spin-flip scattering, resulting in a nearly
temperature independent phase coherence time over a broad
temperature range~\cite{PRBtauphi}. As shown in
Ref.~\cite{PRBtauphi}, this mechanism could explain the apparent
low-temperature saturation of $\tau_{\varphi}$ observed in many
experiments, which caused a controversy in recent years
\cite{MW,Lin}. It was recently proposed that magnetic impurities
also affect the energy exchange rate between electrons~\cite{KG},
which could explain the anomalous interaction rate observed in a
series of experiments~\cite{Pothier97,relaxAg}. A first hint that
this proposal is relevant was the observation of a magnetic field
dependence of the rate~\cite{PRLAnne,TheseAnne}, in a manner
consistent with a theory taking into account the Kondo
effect~\cite{GGAG}. In those experiments, however, the nature and
amount of magnetic impurities were not controlled. Assuming that the
impurities were Mn, the concentrations needed to explain energy
exchange experiments in silver wires were up to two orders of
magnitude larger than the concentrations deduced from
$\tau_{\varphi}$ measurements on similar
samples~\cite{PRLAnne,TheseAnne}. It was proposed that the samples
for energy exchange rate measurements could have been contaminated
during fabrication~\cite{PRLAnne,TheseAnne}. Another hypothesis is
that impurities other than Mn , which affect energy exchange rates
more drastically then phase coherence, were present~\cite{GGHG,
Ursule}. Comparison of these proposals with existing experimental
results is difficult because it requires dealing with more involved
theories (large spin, surface anisotropy, large Kondo temperature),
and pointless because it requires uncontrolled extra parameters. In
order to overcome these difficulties and investigate quantitatively
the mechanism proposed by Ref.~\cite{KG}, we have performed a
comparative experiment described in this Letter, in which we probe
the specific effect of the addition of 0.7~ppm (parts per million)
of Mn atoms on energy exchange rate between electrons. We measured
the temperature dependence of $\tau_{\varphi}$ on the same samples,
accessing interactions in a complementary manner.

The scattering of electrons by magnetic impurities in metals is a
many-body problem known as the Kondo effect: electrons tend to
screen the spin of the impurity, leading to a renormalization of the
scattering rate. The characteristic energy scale for this process is
the Kondo temperature $T_K$. At $T\gtrsim T_K$, screening is
incomplete, and spin-flip scattering takes place, whereas, at $T\ll
T_K$, the impurity and the electrons form a singlet state, leading
to potential scattering only. As far as electron dephasing is
concerned, Kondo effect results in a maximal dephasing rate at
$T_K$~\cite{vanDelft}. Kondo effect also provides a channel for
efficient energy exchange between electrons scattering from the same
magnetic impurity~\cite{KG,Georg,Kroha}. The rate of such a process
depends on the energy of the states of the magnetic impurity, and is
therefore sensitive to magnetic field because of the Zeeman
effect~\cite{GGAG}. The spin states of the magnetic impurities can
furthermore be split in presence of spin-orbit scattering near an
interface~\cite{Zawa}, which also modifies the rate. Further
complication arises when the concentration of magnetic impurities is
so high that the RKKY interaction between magnetic impurities
constrains the spin dynamics~\cite{Vavilov,Schopfer}.

In order to test quantitatively the impact of magnetic impurities on
energy exchange between electrons, we have compared the energy
exchange rate and $\tau_{\varphi}(T)$ in two wires that differ only
by the intentional addition of manganese impurities in one of them,
with concentration low enough so that interactions between Mn
impurities can be neglected~\cite{Vavilov}. To observe specifically
the influence of the Mn impurities, the two samples were fabricated
simultaneously on the same wafer. In a first step, a set of wires
and their contact pads were patterned by e-beam lithography and
evaporation of silver from a nominally 6N-purity source (99.9999\%
Ag from Alfa Aesar$^\circledR$). Mn$^{+}$ ions were implanted at
$70~$kV in half of them, using the ion implanter IRMA at CSNSM
Orsay. The neutralization current from the sample holder to ground
was monitored during the implantation, leading to a direct
measurement of the number of implanted atoms. Monte Carlo
simulations~\cite{TRIM} yield the concentration of Mn atoms that
stop inside the silver wire $c=0.7\pm 0.1$~ppm. In order to measure
the energy exchange between electrons~\cite{Pothier97}, a long and
thin electrode forming a tunnel junction with the middle of the wire
is used as a probe. This electrode was patterned on individual chips
in a second lithography step followed by evaporation of 3.5~nm of
aluminum, oxidation, and evaporation of 16~nm of aluminum. We focus
here on the results obtained on two wires, one without manganese
added (labeled ``bare'' in the following), one with manganese added
(``implanted''). For both samples, the wire length and cross-section
area are $L=40~\mu$m and $\mathcal{S}_e=230~\mathrm{nm}\times
42~\mathrm{nm}.$ The samples were measured in a dilution
refrigerator with base temperature of $20~$mK. The low temperature
wire resistance ($R=55~\Omega )$ was identical for both wires, which
yields the diffusion constant of electrons
$D=0.029~\mathrm{m}^{2}/\mathrm{s}$.

\begin{figure}[hptb]
\begin{center}
\includegraphics[width=2.8in]{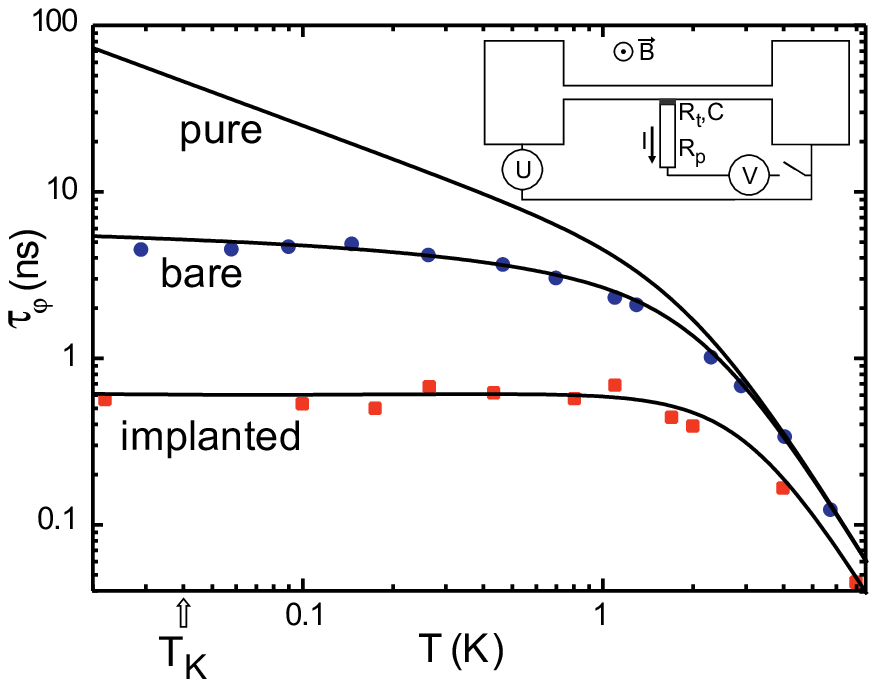}
\caption{(Color online) Symbols: measured phase coherence time in
the two wires. Solid lines: best fits with Eq.~(\ref{tauphieq}),
obtained with $c_b=0.10\pm 0.01$~ppm (bare wire) and $c_i=0.95\pm
0.1$~ppm (implanted wire). The upper line is the prediction without
spin-flip scattering ($c=0$). Inset: layout of the circuit. The
switch is open for magnetoresistance measurements, closed for energy
exchange measurements. \label{tauphi}}
\end{center}
\end{figure}

For each wire, we have first measured the magnetoresistance at
temperatures ranging from 20~mK to 7~K. Following
Ref.~\cite{Aleiner,PRBtauphi}, magnetoresistance curves are fit
using the theory of weak localization, resulting in evaluations
of the phase coherence time $\tau_{\varphi}$. In the bare wire,
it was important to take into account finite length corrections
because $\tau_{\varphi}$ is comparable to the diffusion time
$\tau_D=L^2/D \approx 56~$ns below 1~K~\cite{Gilles}, leading to
a reduction of the predicted magnetoresistance by $\approx 30\%$
below 1~K. Reproducible conductance fluctuations were visible, so
that the uncertainty in the determination of $\tau_{\varphi}$
becomes large below 60~mK in the bare sample. The spin-orbit time
$\tau_\mathrm{so}\approx 8~\rm{ps}$ was extracted from the data
above 1~K. The temperature dependence of $\tau_{\varphi}$ is
shown in Fig.~\ref{tauphi} for both wires. Below 1~K,
$\tau_{\varphi}$ is smaller by nearly one order of magnitude in
the implanted wire than in the bare one. In none of the samples
does $\tau_{\varphi}$ increase as $T^{-2/3}$ when temperature is
lowered, as would be expected if the electron-electron
interaction was the dominant dephasing process (solid line
labeled ``pure'' in Fig.~\ref{tauphi}). The apparent saturation
of $\tau_{\varphi}$ is attributed to the presence of magnetic
impurities~\cite{PRBtauphi}. This effect is quantified by a fit
of the data with a sum of three terms:
\begin{equation}
\frac{1}{\tau_{\varphi}}=\mathcal{A} T^{2/3}+\mathcal{B}
T^3+\gamma_\mathrm{sf}(T), \label{tauphieq}
\end{equation}
with $\mathcal{A}=\frac{1}{\hbar}\left(\frac{\pi k_B^2}{4 \nu_F L
\mathcal{S}_e}\frac{R}{R_K}\right)^{1/3}$ describing Coulomb
interaction~\cite{Aleiner}, $\mathcal{B}$ electron-phonon
interaction~\cite{Lin} and
\begin{equation}
\gamma_\mathrm{sf}(T)=\frac{c}{\pi\hbar\nu_F}\frac{\pi^2
S(S+1)}{\pi^2 S(S+1)+\ln(T/T_K)^2}
\end{equation}
the spin-flip scattering rate, according to Nagaoka-Suhl formula
\cite{PRBtauphi,effetdeBsurtausf}. The density of states in silver
is $\nu_F\approx 1.03\times 10^{47}~$J$^{-1}$m$^{-3}$ (2 spin
states), the resistance quantum $R_K=h/e^2$, and the spin of the
magnetic impurities $S$. Assuming that the only magnetic impurities
present are Mn atoms, with $S=5/2$ and
$T_K=40~\mathrm{mK}$~\cite{Suhl} and that $\mathcal{A}$ is fixed at
its theoretical value $\mathcal{A}=0.19~\textrm{ns}^{-1}K^{-2/3}$,
the best fits are obtained for $c_b=0.10\pm 0.01$~ppm and
$\mathcal{B}_b\approx 3.7\times 10^{-2}$~ns$^{-1}$K$^{-3}$ for the
bare wire, and $c_i=0.95\pm 0.1$~ppm, $\mathcal{B}_i\approx
5.5\times 10^{-2}$~ns$^{-1}$K$^{-3}$ for the implanted
one~\cite{phonons}. The difference between the implanted and bare
samples, $c_i-c_b=0.85\pm0.1$~ppm, is in reasonable agreement with
the estimated amount of implanted ions. The value of $c_b$ is
significantly larger than found in previous
experiments~\cite{PRBtauphi}, indicating a lesser quality of the
source material or a slight contamination during fabrication.

We have then measured the energy exchange rate between electrons and
its dependence on magnetic field $B$ on the same two wires. The
principle of the experiment is to drive electrons out-of-equilibrium
with a bias voltage $U\gg k_BT/e$. The distribution function $f(E)$
of the electrons in the middle of the wire depends crucially on
energy exchange between electrons \cite{Pothier97}. The differential
conductance $\de I/\de V(V)$ of the tunnel junction between the wire
and the probe electrode (inset of Fig.~\ref{tauphi}, switch closed;
see also Ref. \cite{PRLAnne}) is a convolution product of $f(E)$
with a function $q(E)$ describing inelastic
tunneling~\cite{PRLAnne}:
\begin{equation}
R_t \frac{\de I}{\de V}(V)=1-\int f(E)q(eV-E)\de E \label{convol}
\end{equation}
where $R_t$ is the resistance of the tunnel junction. The
information on $f(E)$ is therefore contained in $\de I/\de V(V)$
\emph{via} the $q$ function. The experiment is performed at $B\geq
0.3~T$, and the aluminum probe electrode is in its normal state. The
$q$ function is obtained from $\de I/\de V(V)$ at $U=0$, where
$f(E)$ is a Fermi function.  In this situation, $\de I/\de V(V)$
displays a sharp minimum at zero voltage (sometimes called ``zero
bias anomaly''), due to dynamical Coulomb blockade of
tunneling~\cite{GrabertDevoret}. The environmental impedance
responsible for Coulomb blockade is the resistance $R_p$ of the
probe electrode. The conductance is reduced at $V=0$ by a factor
$0.78$ in the bare sample and $0.62$ in the implanted one. A slight
($3\%$ at most), unexpected dependence on $B$ of $\de I/\de V(V)$
was observed on the implanted sample. In practice, we therefore
derived a $q$ function at each value of $B$ from $\de I/\de V(V)$
taken at $U=0$. Fits of $\de I/\de V(V)$~\cite{JoyezEsteve} give the
resistance of the environment $R_p=0.95~\mathrm{k}\Omega$
(respectively, $1.3~\mathrm{k}\Omega$), the capacitance of the
tunnel junction $C=4.4~\mathrm{fF}$ ($\approx 0.7~\mathrm{fF}$), the
tunnel resistance $R_t=16.5~\mathrm{k}\Omega$
($96.9~\mathrm{k}\Omega$) and the temperature $T_0=45~\mathrm{mK}$
for the bare (implanted) sample. The differences in those parameters
are essentially due to geometry, and do not interfere with the
measurement of energy exchange between electrons in the wires. When
electrons are driven out-of equilibrium ($U \neq 0$), $f(E)$ is not
a Fermi function any longer. In the absence of energy exchange,
$f(E)$ presents two steps at $E=-eU$ and $E=0$, resulting in a
splitting of the dip in $\de I/\de V(V)$ into two dips. In the
opposite limit of very high energy exchange rate, $f(E)$ approaches
a Fermi function at a temperature $T \approx \frac{\sqrt{3}}{2\pi}
\frac{eU}{k_B}$, and $\de I/\de V(V)$ presents a broad
dip~\cite{PRLAnne}.

\begin{figure}[hptb]
\begin{center}
\includegraphics[width=8.6cm]{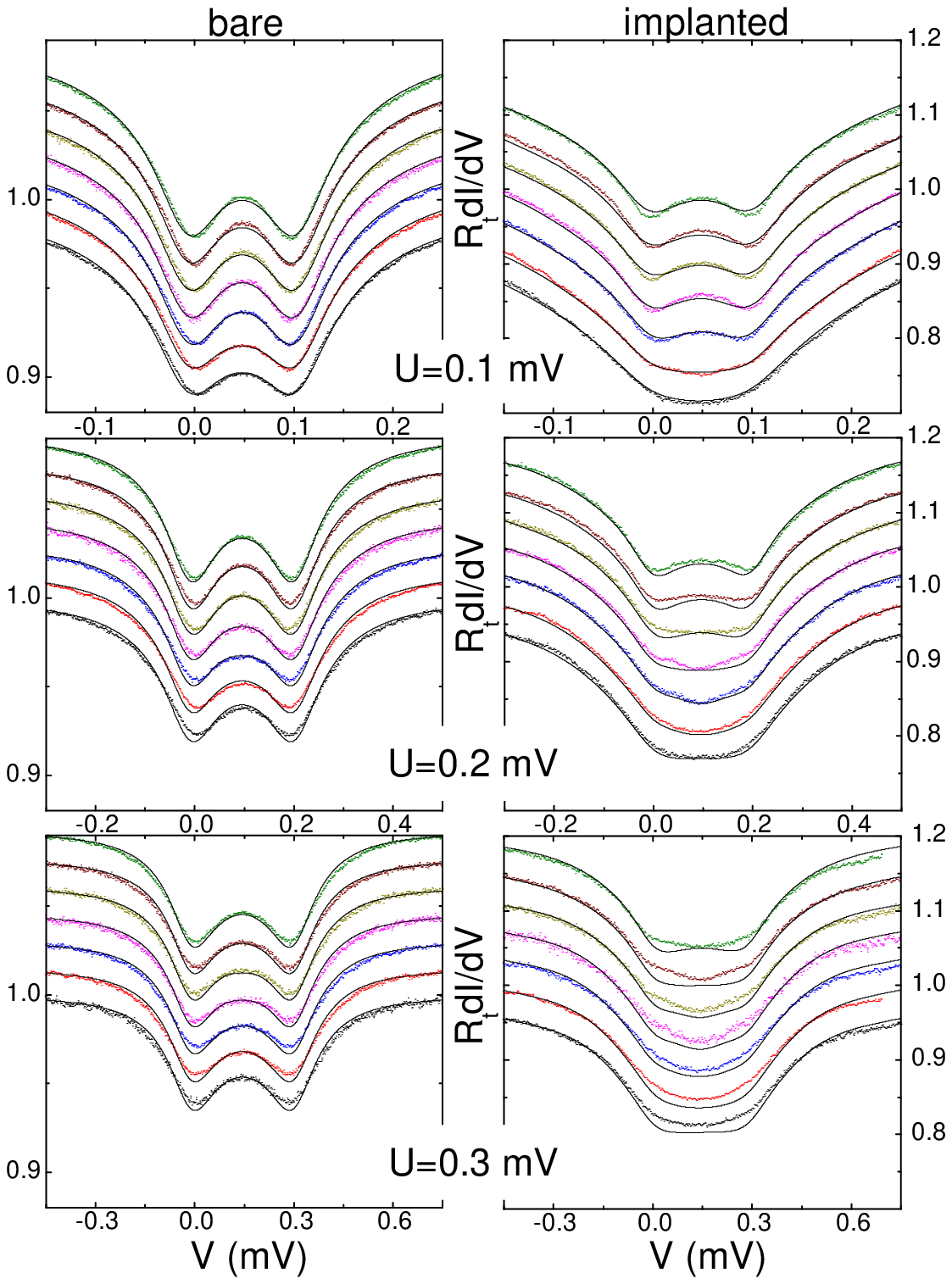}
\caption{(Color online) Differential conductance $\de I/\de V(V)$ of
the tunnel junction (see inset of Fig.~\ref{tauphi}) for the bare
(left) and implanted (right) wires, for $U=0.1$~mV, 0.2~mV and
0.3~mV (top to bottom panels), and for $B=0.3$ to $2.1~$T by steps
of $0.3~$T (bottom to top in each panel). The curves were shifted
vertically for clarity. Symbols: experiment. Solid lines:
calculations using $c_b=0.1$~ppm, $c_i=0.95$~ppm and $\kappa_
{ee}=0.05~\mathrm{ns}^{-1}\mathrm{meV}^{-1/2}$.  \label{data}}
\end{center}
\end{figure}

In Fig.~\ref{data}, we show the measured $\de I/\de V(V)$
characteristics of the tunnel junctions on the bare and implanted
wires, for $U=0.1,$ 0.2 and 0.3~mV, and for $B$ ranging from 0.3~T
to 2.1~T by steps of 0.3~T. At $B=0.3~$T, the measurements on the
bare sample show two clear dips at $V=0$ and $V=U$, whereas the
measurements on the implanted sample show a single, broad dip around
$V=U/2$. The addition of 0.7~ppm of Mn has therefore significantly
increased the energy exchange rate between electrons, resulting in a
strong energy redistribution during the diffusion time
$\tau_D=56~$ns. At $B=2.1~$T, the broad dip found in the implanted
sample has split into two dips for $U=0.1$ and 0.2~mV, indicating
that the energy exchange rate due to the Mn impurities is now
smaller than $1/\tau_D$.

The coupling between electrons and magnetic impurities can be
described by an exchange Hamiltonian, characterized by a coupling
constant $J$. At zero magnetic field, this description leads to
energy exchange in second order perturbation theory, as described in
Ref.~\cite{KG}. At finite magnetic field, the spin states of the
impurities are split by the Zeeman energy $E_Z=g \mu_B B$. The
energy $E_Z$ can then be exchanged at the lowest order in
perturbation theory between electrons and impurities. This approach
is sufficient to understand qualitatively the magnetic field
behavior: the rate of interaction decays rapidly when $E_Z>eU$,
because very few electrons can excite the impurities. The magnetic
fields $eU/g\mu_B$ (using $g=2$ for Mn) are 0.86, 1.7 and 2.6~T for
$U=0.1,$ 0.2 and 0.3~mV, which correspond in the implanted wire to
the fields at which the curvature of $\de I/\de V(V)$ near $V=U/2$
changes sign. In the bare sample, the double dip also gets sharper
when $B$ is increased. This is an indication that, as inferred from
$\tau_{\varphi}(T)$ measurements, this sample also contained some
magnetic impurities. However, the corresponding energy exchange rate
is always smaller than $1/\tau_D$, and $\de I/\de V(V)$ displays a
double dip.

In order to compare quantitatively the measurements with theory, the
renormalization of the coupling constant $J$ by Kondo effect needs
to be considered. Very roughly, this renormalization amounts
to~\cite{KG} $J_{\mathrm{eff}}/J\approx [\nu_F J\ln(eU/k_B
T_K)]^{-1}\approx 3$. More precisely, $J_{\mathrm{eff}}$ depends on
the distribution function $f(E)$, and only the full theory of
Ref.~\cite{GGAG} is able to quantify this effect and to treat the
exchange Hamiltonian at all orders on the same footing. We have
therefore solved the Boltzmann equation for $f(E)$
self-consistently, taking into account Coulomb interaction,
electron-phonon interaction~\cite{TheseFred} and the effect of
magnetic impurities in a magnetic field following the full theory of
Ref.~\cite{GGAG}. The concentration of magnetic impurities and the
electron-phonon coupling were fixed at the values determined from
the fit of $\tau_{\varphi}(T)$~\cite{TheseFred}. We used
$T_K=40~\mathrm{mK}$~\cite{Suhl} and $g=2.0$~\cite{Brodale}. Note
that theory assumes $S=1/2$ whereas $S=5/2$ for Mn atoms, but it is
not expected that this difference has a large influence on energy
exchange~\cite{GGHG}. The intensity of Coulomb interaction alone
could not be determined accurately from $\tau_{\varphi}(T)$, and
since it was found that theory underestimates the intensity
$\kappa_{ee}$ of Coulomb interaction~\cite{Coulomb}, $\kappa_{ee}$
was used as a free parameter, common to both samples. A slight
increase in temperature of the contact pads of the wire with $U$
(0.76~K/mV) was taken into account~\cite{TheseFred}. We also
included in the calculation a slight heating of the electrons in the
probe electrode at the junction interface, due to the fact that
$R_p$ is not negligible compared to $R_t$. The corresponding
temperature $T_p(U,V)$ of the electrons in the probe electrode is
$T_p\approx 0.34~$K in the bare and 0.16~K in the implanted sample
at the dips ($V=0$ or $U$), at $U=0.3~$mV where $T_p$ is expected to
be the largest. The differential conductance $\de I/\de V(V)$ was
then computed using Eq.~(\ref{convol}). The resulting curves are
displayed as solid lines in Fig.~\ref{data}. The best agreement
between theory and all the data was found for $\kappa_
{ee}=0.05~\mathrm{ns}^{-1}\mathrm{meV}^{-1/2}$. This value is larger
than the prediction
$\kappa^{\mathrm{AAK}}_{ee}=0.016~\mathrm{ns}^{-1}\mathrm{meV}^{-1/2}$~\cite{AAK},
as was repeatedly found in previous experiments~\cite{Coulomb}. A
good overall agreement is found for both data sets, but some
discrepancy appears for the implanted sample at $U=0.3~$mV. We
evaluated the sensitivity of the fits of the data on the implanted
wire to the concentration $c_i$ of the impurities, and found that
the best agreement is obtained at $c_i=0.9\pm 0.3$~ppm, in good
agreement with the value 0.95~ppm deduced from the data of
Fig.~\ref{tauphi}.

In conclusion, in this comparative experiment, the observed effect
of well-identified magnetic impurities on energy exchange is found
to be in good quantitative agreement with the theory of
Ref.~\cite{GGAG}, the concentration of impurities being fixed to the
value deduced from the temperature dependence of the phase coherence
time, which is also compatible with the expected value from
implantation. This well-controlled experiment shows that the
interaction mediated by dilute, low Kondo temperature magnetic
impurities is well understood. However it remains that, in this
experiment as in all previous ones, Coulomb interaction seems to be
more efficient for energy exchange than predicted~\cite{Coulomb}.
Open questions remain also on the contribution of Kondo effect to
dephasing and energy exchange at energies below $T_K$
\cite{vanDelft}, on the effect of the interactions between
impurities at larger concentrations \cite{Vavilov,Schopfer} and on
finite size effects \cite{Ursule}.

\begin{acknowledgments}
This work was supported in part by EU Network DIENOW. We acknowledge
the assistance of S.~Gautrot, O. Kaitasov and J.~Chaumont at the
CSNSM in Orsay University, who performed the ion implantation. We
gratefully acknowledge discussions with F.~Pierre, H. Gra\-bert,
G.~G\"oppert, A.~Zawadowski and H.~Bouchiat.
\end{acknowledgments}


\begin{thebibliography}{}
\bibitem{effetdeBsurtausf} The width of the weak localization dip in the magnetoresistance, $\approx 10~$mT, was small enough to neglect the
variation of the spin-flip rate with magnetic field.
\bibitem{phonons} The difference between the $\mathcal{B}$ parameters is not understood, but it only affects the dependence of $\tau_{\varphi}(T)$ above 2~K, whereas the impurity
concentrations $c_i$ and $c_b$ are determined by low-T behavior.
\end{thebibliography}
\end{document}